\documentclass[9pt,twocolumn,twoside]{pnas-new}
\usepackage{dblfloatfix}
\usepackage{float}
\newcommand{\be}{\begin{equation}}
\newcommand{\ee}{\end{equation}}

\templatetype{pnasresearcharticle}

\title{Interrogating the \emph{E. coli} cell cycle by cell dimension perturbations}

\author[a,b]{Hai Zheng}
\author[c]{Po-Yi Ho}
\author[a]{Meiling Jiang}
\author[d]{Bin Tang}
\author[a,b]{Weirong Liu}
\author[a]{Dengjin Li}
\author[e]{Xuefeng Yu}
\author[f]{Nancy Kleckner}
\author[c,1]{Ariel Amir}
\author[a,b,1]{Chenli Liu}

\affil[a]{Center for Synthetic Biology Engineering Research, Shenzhen Institutes of Advanced Technology, Chinese Academy of Sciences, Shenzhen 518055, P. R. China}
\affil[b]{University of Chinese Academy of Sciences, Beijing, 100049, P. R. China}
\affil[c]{School of Engineering and Applied Sciences, Harvard University, Cambridge, MA 02138, USA}
\affil[d]{Department of Materials Science and Engineering, Southern University of Science and Technology, Shenzhen 518055, P. R. China}
\affil[e]{Institute of Biomedicine and Biotechnology, Shenzhen Institutes of Advanced Technology, Chinese Academy of Sciences, Shenzhen 518055, P. R. China}
\affil[f]{Department of Molecular and Cellular Biology, Harvard University, Cambridge, MA 02138, USA}

\leadauthor{Zheng}

\significancestatement{How bacteria regulate cell division to achieve cell size homeostasis, with concomitant coordination of DNA replication, is a fundamental question. Currently, there exist several competing models for cell cycle regulation in \emph{E. coli}. We performed experiments where we systematically perturbed cell dimensions, and found that average cell volume scales exponentially with the product of the growth rate and the time from initiation of DNA replication to the corresponding cell division. Our data support a model in which cells initiate replication on average at a constant volume per origin, and divide a constant time thereafter.}

\authorcontributions{HZ, MJ, BT, WL, DL, XY, and CL performed the experiments. PH, NK, AA, and CL analyzed the data and wrote the manuscript. CL designed the research.}
\authordeclaration{The authors declare no conflicts of interests.}
\correspondingauthor{\textsuperscript{1}To whom correspondence should be addressed. E-mail: AA - arielamir@seas.harvard.edu, CL - cl.liu@siat.ac.cn}


\begin{abstract}
Bacteria tightly regulate and coordinate the various events in their cell cycles to duplicate themselves accurately and to control their cell sizes. Growth of \emph{Escherichia coli}, in particular, follows a relation known as Schaechter's growth law. This law says that the average cell volume scales exponentially with growth rate, with a scaling exponent equal to the time from initiation of a round of DNA replication to the cell division at which the corresponding sister chromosomes segregate. Here, we sought to test the robustness of the growth law to systematic perturbations in cell dimensions as achieved by varying the expression levels of \emph{mreB} and \emph{ftsZ}. We found that decreasing \emph{mreB} level resulted in increased cell width, with little change in cell length, whereas decreasing \emph{ftsZ} level resulted in increased cell length. Furthermore, the time from replication termination to cell division increased with the perturbed dimension in both cases. Moreover, the growth law remained valid over a range of growth conditions and dimension perturbations. The growth law can be quantitatively interpreted as a consequence of a tight coupling of cell division to replication initiation. Thus, its robustness to perturbations in cell dimensions strongly supports models in which the timing of replication initiation governs that of cell division, and cell volume is the key phenomenological variable governing the timing of replication initiation. These conclusions are discussed in the context of our recently proposed ``adder-per-origin'' model, in which cells add a constant volume per origin between initiations and divide a constant time after initiation.
\end{abstract}

\dates{This manuscript was compiled on \today}
\doi{\url{www.pnas.org/cgi/doi/10.1073/pnas.XXXXXXXXXX}}

\begin{document}

\verticaladjustment{-2pt}

\maketitle
\thispagestyle{firststyle}
\ifthenelse{\boolean{shortarticle}}{\ifthenelse{\boolean{singlecolumn}}{\abscontentformatted}{\abscontent}}{}

\dropcap{B}acteria can regulate tightly and coordinate the various events in their cell cycles in order to accurately duplicate their genomes and to homeostatically regulate their cell sizes. This is a particular challenge under fast growth conditions where cells are undergoing multiple concurrent rounds of DNA replication. Despite much progress, we still have an incomplete understanding of the processes that coordinate DNA replication, cell growth, and cell division. This lack of understanding is manifested, for instance, in discrepancies among various recent studies that propose different models for control of cell division in the bacterium \emph{Escherichia coli}.

One class of models suggests that cell division is triggered by the accumulation of a constant size (e.g. volume, length, or surface-area) between birth and division \cite{JW,sattar,theriot}. Such models are supported by experiments measuring correlations between cell size at birth and cell size at division, which showed that, when averaged over all cells of a given birth size $v_B$, cell size at division $v_D$ approximately follows:

\be v_D= v_B + v_0 ,\ee
where the constant $v_0$ sets the average cell size at birth. This is known as the ``incremental" or ``adder" model, and cells following this behavior are said to exhibit ``adder correlations'' \cite{AmirPRL,JW,sattar,xavier,ilya,kennard,theriot}. Importantly, these models postulate that cell division is governed by a phenomenological size variable, with no explicit reference to DNA replication.

A second class of models for control of cell division postulates that cell division is governed by the process of DNA replication, which can be described as follows. The time from a replication initiation event to the cell division that segregates the corresponding sister chromosomes can be split into the $C$ period, from initiation to termination of replication, and the $D$ period, from termination of replication to cell division \cite{baclifeseq, levin_review}. Both the $C$ and $D$ periods remain constant at approximately $40$ and $20$ minutes, respectively, for cells grown in various growth media supporting a range of doubling times between $20$ and $60$ minutes \cite{cooperhelmstetter, elf}. We will refer to growth rates within this range as \emph{fast}. All experiments described here are carried out under such fast growth conditions. Note that $C+D$ is approximately $60$ minutes and larger than the time between divisions at fast growth. This situation is achieved by the occurrence of multiple ongoing rounds of replication. That is, under these conditions, a cell initiates a round of replication simultaneously at multiple origins that ultimately give rise to the chromsomes of their grand- or even great-grand- daughters \cite{SkarstadBoye}. Extending the basic definition of the $C$ and $D$ periods, Cooper and Helmstetter specifically proposed that an initiation event triggers a division after a time $C+D$, thereby ensuring that cells divide only after the completion of a round of DNA replication \cite{cooperhelmstetter, baclifeseq}. This Cooper-Helmstetter (CH) formulation, hereafter the CH model, belongs to the second class of models for control of cell division.

\begin{figure}[t]
\centering
\includegraphics[width=2.5in]{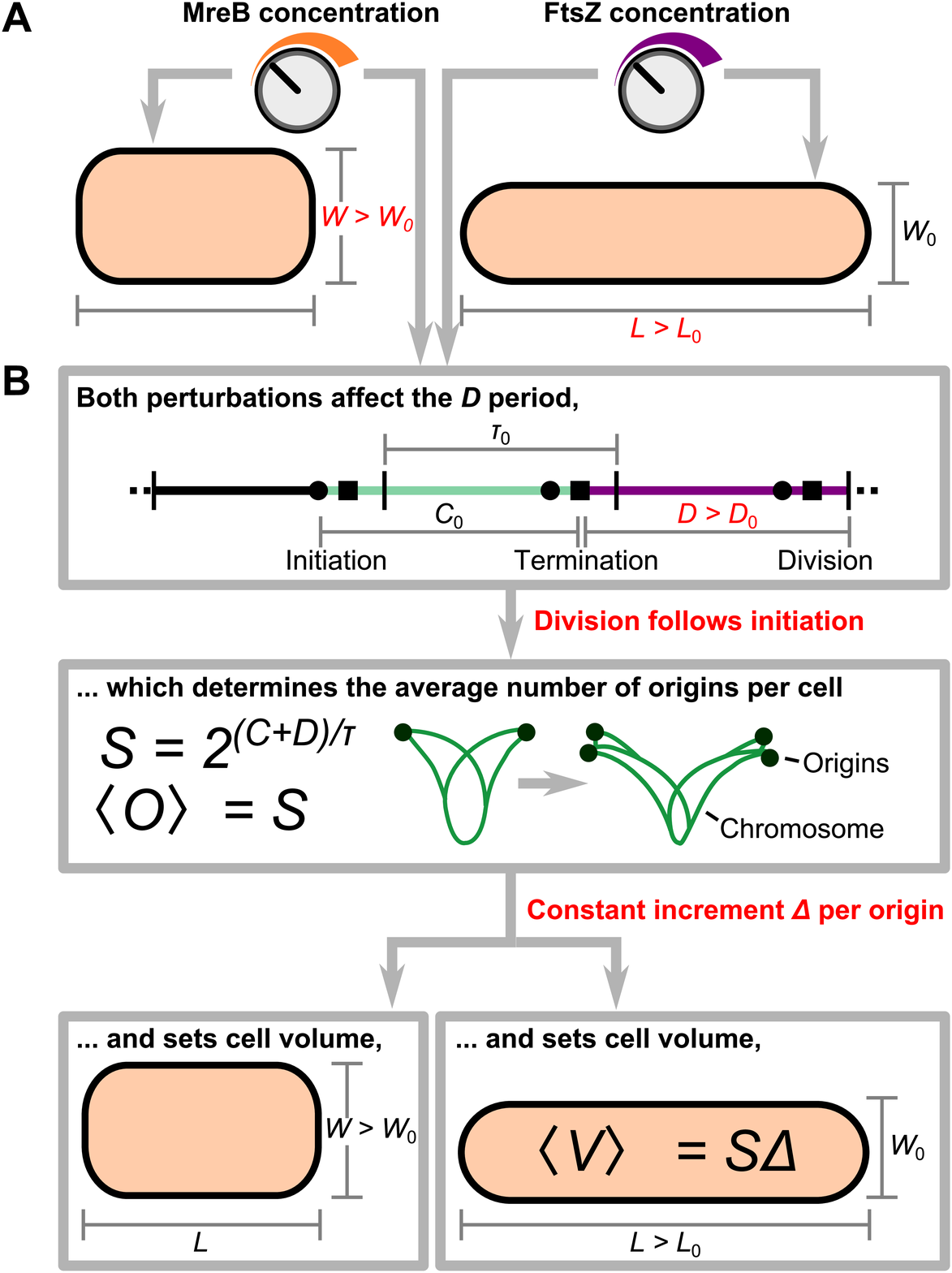}
\caption{(A) Schematic illustration of the experiment. MreB and FtsZ are involved in cell wall synthesis and septum formation, respectively. Using \emph{mreB}- or \emph{ftsZ}-titratable strains, we are able to tune their expression levels continuously, and perturb cell dimensions. In both experiments, the $D$ period increased with cell width and length. The $C$ period and doubling time $\tau$ remained constant. (B) Schematic illustration of our model. The perturbed $D$ period sets the average number of origins per cell, which is equal to the scaling factor $S$ because replication initiation triggers cell division. The average number of origins per cell then sets the average cell volume following the growth law. For titrated \emph{mreB} levels, cell volume changes manifested mostly as cell width changes. For titrated \emph{ftsZ} levels, cell length changed instead, because FtsZ did not affect cell width.}
\label{Fig1}
\end{figure}

The CH model is supported by the phenomenon of rate maintenance \cite{shiftup}: after a change from one growth medium to a richer one (a shift-up), cells continue to divide at the rate associated with the poorer medium for a period of 60 minutes. According to the CH model, in which division is triggered by initiation, all cells that have already initiated replication before a shift-up will have also already committed to their ensuing divisions, and thus the rate of division will remain unchanged for a time $C+D$ following a shift-up.

The same value of 60 minutes had also emerged in a seemingly different context a decade earlier, in the seminal study of Schaechter \emph{et al.} \cite{schaechter}. In their work, cell volumes, averaged over an exponentially growing population, were measured for culture growing under dozens of different growth media supporting fast growth. It was found that average cell volume was well described by an exponential relation with growth rate $V = \Delta e^{\lambda T}$, where $V$ is the average cell volume, $\Delta$ is a constant with dimensions of volume, $\lambda$ is the growth rate, $\tau=\log(2)/\lambda$ is the doubling time, and $T \approx 60$ mins.

Donachie showed that this exponential scaling of average cell volume with growth rate can also be explained by the CH model if it is further assumed that cells initiate replication on average at a constant volume $\Delta_I$ per origin of replication at initiation \cite{donachie68}. Because cells grow exponentially at the single-cell level \cite{manalis}, cells will then divide on average at a volume $\Delta_I$ per origin times a scaling factor $S = 2^{(C+D)/\tau}$. The average cell volume then follows

\be V = \Delta 2^{(C+D)/\tau} = S \Delta ,\label{growthlaw} \ee
with $\Delta = \log(2)\Delta_I$, because the cell volume averaged over an exponentially growing population is the average cell volume at birth times $2\log(2)$ \cite{powell}. In Schaechter's experiments, the $C$ and $D$ periods were approximately constant, giving rise to the exponential scaling observed. However, the derivation for Eq. \ref{growthlaw} holds regardless of the values of the $C$ and $D$ periods, and in cases where they are not constant, average cell volume is not expected to scale exponentially with growth rate. Eq. \ref{growthlaw} is known as Schaechter's growth law, but will be referred to simply as the growth law for the rest of this manuscript.

Recent single-cell analyses found that cells indeed initiate replication on average at a constant volume per origin or per some locus close to the origin \cite{elf}. While further experiments are required, the fact that introduction of an origin onto a plasmid does not affect cell cycle timings or cell size suggests that the latter possibility is correct \cite{leonard,twoori,elf} (\emph{SI}). Below, for simplicity, we will use the phrase ``per origin,'' while keeping this complexity in mind.

\begin{figure}[b]
\centering
\includegraphics[width=3in]{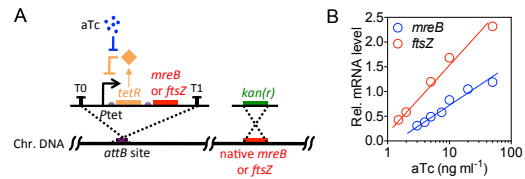}
\caption{Titratable \emph{mreB} or \emph{ftsZ} expression. (A) The genetic circuit of the \emph{mreB}- or \emph{ftsZ}-titratable strains. The expression of \emph{mreB} or \emph{ftsZ} is under the control of a P$_{tet}$-\emph{tetR} feedback loop and the native \emph{mreB} or \emph{ftsZ} was seamlessly replaced with a kanamycin resistance gene. (B) Relative \emph{mreB} and \emph{ftsZ} mRNA level in the titratable strains in bulk culture containing various concentrations of aTc (3-50 ng ml$^{-1}$).}
\label{Fig2a}
\end{figure}

Clearly, the two classes of models for control of cell division differ fundamentally. In the first class, division depends only on the accumulation of size from birth, and DNA replication plays no explicit role. In the second class, division is downstream of the preceding initiation of DNA replication. Importantly, also, the experiments leading to the first class of models defined cell size differences by measuring cell length. Since for a constant growth environment the widths of bacterial cells are very narrowly distributed with a coefficient of variation (CV) less than $0.05$ \cite{sattar}, these analyses cannot distinguish whether cell size in a given environment is set by a constant volume, surface-area, or length. This ambiguity raises the question of what is the key phenomenological variable governing cell cycle progression.

\begin{figure*}[b]
\centering
\includegraphics[width=7in]{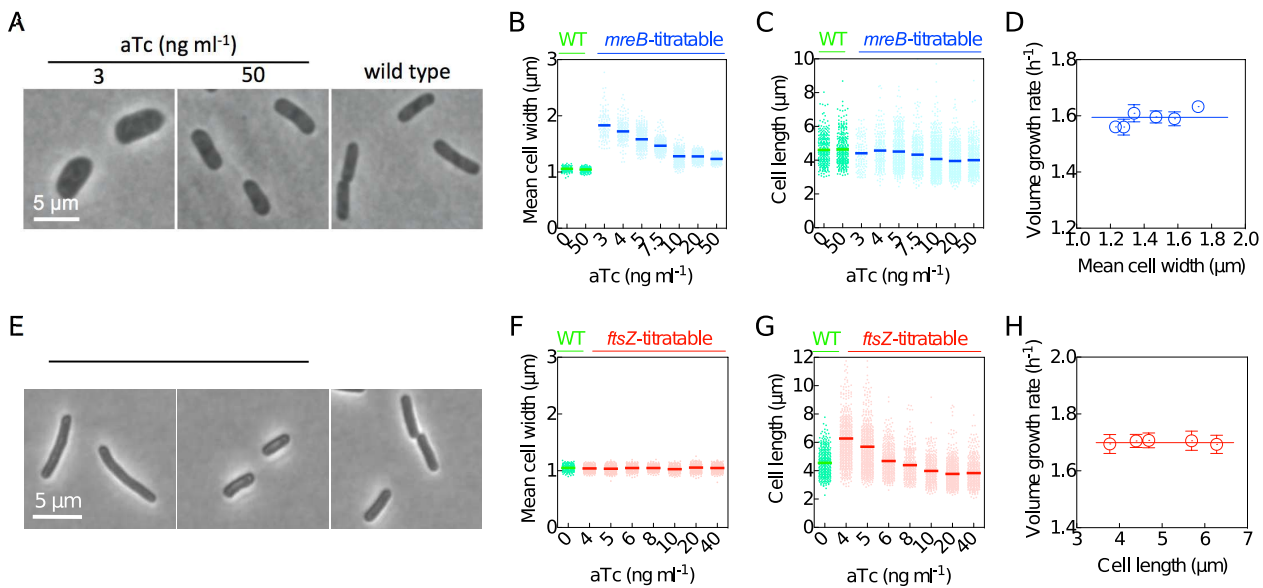}
\caption{Titratable \emph{mreB} or \emph{ftsZ} expression to systematically perturb cell width or cell length, respectively, without affecting the volume growth rates. (C) Representative phase contrast images of the \emph{mreB}-titratable and wild type strains. (D and E) Scatter plot presents the average, mean cell width (averaged along the long axis of a cell) (D) or average cell length (E) of the individual wild type (WT) or \emph{mreB}-titratable cells. (F) Volume growth rates in bulk culture versus mean cell width for \emph{mreB}-titratable cells. (G, H, I, and J) The same as (C, D, E, and F) but for the \emph{ftsZ}-titratable strain. Error bars represent the SEM of three replicates.}
\label{Fig2}
\end{figure*}

Here, we sought to test these models by perturbing cell dimensions in \emph{E. coli}, and assaying the effects of those perturbations on both replication events and cell division. In our study, shape perturbations were achieved by systematically varying expression levels of the protein MreB, an actin homologue involved in cell wall synthesis, and the protein FtsZ, a tubulin homologue involved in the formation of the division septum \cite{garner,nhill}. Our approach is indicated schematically in Fig. \ref{Fig1}A. It extends and complements Schaechter's experiments, in which growth rate was perturbed, and is reminiscent of the work of Harris \emph{et al}. \cite{theriot}, in which cell dimensions were perturbed, but with the important addition to both studies that we also measured the cell cycle periods $C$ and $D$ because they play an important role in the CH model. For this paper, we define division as completion of septation.

\section*{Results}

\subsection{Decreased \emph{mreB} level resulted in increased cell width, with little change in cell length}
Cell length and cell width are two major characteristics of a rod-shaped cell. As a cell grows, cell length increases exponentially, while cell width remains constant. How cell width is determined and maintained is largely unknown. However, several mutations of \emph{mreB} were reported to result in altered cell dimensions \cite{wangMreB,garner,sven}, thus raising the possibility that alterations in \emph{mreB} expression level would alter cell width. To continuously and systematically vary cell width, we constructed a strain in which the level of \emph{mreB} could be experimentally controlled. We employed a system in which a \emph{P}$_{tet}$-\emph{tetR} feedback loop triggered \emph{mreB} expression (Fig. \ref{Fig2a}A, and \emph{SI}). The modulated copy of \emph{mreB} was the sole version of the gene in the genome as the native copy of \emph{mreB} was replaced by a kanamycin-resistance gene. In this construct, expression of \emph{mreB} could be tightly controlled by adjusting the concentration of an appropriate inducer of \emph{P}$_{tet}$-\emph{tetR}, anhydrotetracycline (aTc).

We found that cell size increased with decreasing inducer concentration until, at very low \emph{mreB} levels, the cells eventually lysed (at an aTc concentration below 1 ng ml$^{-1}$ in Rich Defined Medium (RDM) + glucose). Above this minimum threshold, the expression level of \emph{mreB} varied linearly with the concentration of inducer (Fig. \ref{Fig2a}B). We also found that within a certain range of \emph{mreB} expression levels, the volume growth rates, or OD$_{600}$ doubling rates (Fig. S1), of the titratable strain remained approximately constant (CV of $0.03$) (Fig. \ref{Fig2}D). Together, these results suggest that the titratable system is suitable for characterizing the functions of MreB in a quantitative manner with no need to consider complications due to differing growth rates.

\begin{figure}[t]
\centering
\includegraphics[width=3in]{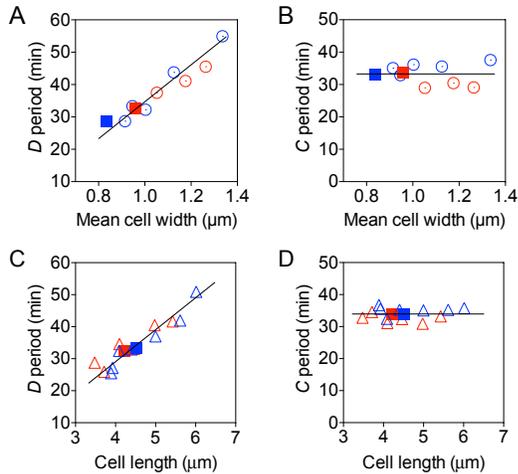}
\caption{Changes in the cell cycle parameters as cell dimensions are perturbed. (A) The $D$ period increased monotonically with cell width in \emph{mreB}-titratable strains. The line is the best linear fit. (B) The $C$ period remained approximately constant as cell width changed in response to titrated \emph{mreB} expression levels. The line is the mean value of $C$ averaged over \emph{mreB} expression levels. (C) The $D$ period increased monotonically with cell length in \emph{ftsZ}-titratable strains. (D) The $C$ period remained approximately constant as cell length changed in response to titrated \emph{ftsZ} expression levels. The circle, triangle, and square indicates \emph{mreB}-titratable, \emph{ftsZ}-titratable, and wild type strains, respectively. Different colors denote growth media: red is RDM+glucose, blue RDM+glycerol. The SEMs of three replicates were smaller than the size of the symbols.}
\label{Fig3}
\end{figure}

We next measured cell dimensions by phase contrast microscopy (Fig. \ref{Fig2}A), and found that with decreasing \emph{mreB} expression level, the cell width increased (Fig. \ref{Fig2}B). The cell length changed slightly, which we discuss below (Fig. \ref{Fig2}C, \emph{SI}). Although the software used for image processing is designed to define bacterial cell dimensions to subpixel precision \cite{microbetracker}, we wanted to verify our results using an independent method. We showed that the OD per cell is linearly correlated with the cell volume in $\mu m^3$ (Fig. S1). We also verified that neither the presence of inducer nor the presence of the genetic circuit construct has any effect on wild type cells within the ranges of inducer concentrations studied here (Fig. \ref{Fig2}BC, and \emph{SI}). Similar results were obtained in all other growth media, supporting various fast growth rates, tested in this study (Fig. S2). All of our experiments, and thus the resultant conclusions, concern fast growth conditions as defined above.

\subsection{Cell width maintenance by cell wall stiffness}
Lastly, we characterized in detail the morphological properties of the \emph{mreB} titrated cells using scanning electron microscopy. Here, wider cells showed a slightly flattened, dumpling-like morphology (Fig. S3A). It is known that MreB is involved in bacterial cell wall synthesis \cite{wangMreB,garner,sven}. Thus it seemed possible that the cell wall might have become softer in cells expressing low levels of MreB. Due to the difficulty in directly measuring cell wall elasticity, we instead measured the effective cellular stiffness (ECS) of the \emph{mreB} titrated cells using atomic force microscopy. As expected, the ECS significantly decreased as cell width increased (Fig. S3B). This result raises the possibility that increased cell width reflects the force balance between turgor pressure and the tensile resistance of the cell wall.

\subsection{Decreased \emph{ftsZ} level resulted in increased cell length, and no change in cell width}
We constructed and characterized, using the same method as above, an \emph{ftsZ}-titratable strain to allow perturbation of cell length (\emph{SI}). We found, in agreement with previous work \cite{nhill}, that cell length increased with decreased \emph{ftsZ} expression levels, but that both cell width and growth rate remained relatively constant within the ranges of inducer concentration studied here (CV of $0.01$ and $0.03$ respectively, across different experiments, Fig. \ref{Fig2}EFGH).

\subsection{Correlations between perturbed cell dimensions and cell cycle timings}
We next investigated how the \emph{mreB} and \emph{ftsZ} expression levels affect the $C$ and $D$ periods. For a bulk culture in steady-state exponential growth, the CH model predicts that the average number of origins per cell, $\langle O \rangle$, scales exponentially with growth rate \cite{bremer96,ho}. This is because in steady-state exponential growth, the number of cells and the total number of origins in the population must both grow exponentially at the same rate. However, because the CH model postulates that division only occurs after a time $C+D$ following the corresponding initiation, the total number of origins will be larger than the number of cells by the scaling factor $S$, defined above. Therefore, $\langle O \rangle = S$. This relation holds regardless of the value of $C+D$.

An expression for the average number of copies $X$ of a gene per cell as a function of the location $m$ of the gene along the chromosome ($m=0$ for \emph{oriC} and $m=1$ for \emph{terC}) can be derived similarly. Under the assumption that replication forks travel at a constant speed, the expression is $X=2^{(C(1-m)+D)/\tau}$ \cite{bremer96}. We used this relation to extract the lengths of the $C$ and $D$ periods from \emph{q}PCR data of the copy numbers of different chromosome loci (including \emph{oriC}, \emph{terC} and a series of different loci between them; \emph{SI}). The $C$ period was also measured independently by the methods in Refs. \cite{bremer1, bremer2} which gave values consistent with those obtained by the \emph{q}PCR method (\emph{SI}). We also measured $\langle O \rangle$ using replication-run-out experiments (\emph{SI}).

These analyses revealed that, over the analyzed ranges in the levels of both \emph{mreB} and \emph{ftsZ} expression, the $C$ period remained unchanged (CV of $0.09$ and $0.05$ for \emph{mreB}- and \emph{ftsZ}-titratable strains, respectively), while the $D$ period increased with increasing cell width or length, respectively (Fig. \ref{Fig3}). Note that the relation between the $D$ period and cell length predicted by our model below is not linear, but appears approximately linear given the particular values of the relevant parameters under the conditions of these experiments.

\subsection{The growth law holds in face of perturbations to cell dimensions}
We find that the growth law, Eq. (\ref{growthlaw}), holds in our experiments, both across a range of growth media (Fig. \ref{Fig3}) and across the two titratable perturbations that drastically affected cell width and cell length (Fig. \ref{Fig4}A). The best fit proportionality constant is $\Delta = 0.55 \pm 0.04 \mu$m$^3$. The plus-minus indicates the $95\%$ confidence interval of the fit. From the derivation of the growth law in the Introduction, we find the average cell size per origin at initiation to be $\Delta_I = 0.79 \pm 0.06 \mu$m$^3$. In contrast, the average cell area, cell length, and cell width are not proportional to the scaling factor $S$ (Fig. \ref{Fig4}B, Fig. S4).

We further tested the validity of a constant $\Delta$ by fixing $\Delta$ to $0.55 \mu m^3$ and calculating the ratio of $\log_2(V/\Delta)$ and $\tau^{-1}$, which is equal to $C+D$ according to Eq. (\ref{growthlaw}) (Fig. \ref{Fig5}A). We found that the values of $C+D$ obtained in this way agree well with independent measurements in both the titratable, as well as wild type, strains  (Fig. \ref{Fig5}B).

\section*{Discussion}
Here, we perturbed cell dimensions and then observed the effects of these perturbations on the cell cycle in order to interrogate the mechanism of cell cycle regulation in \emph{E. coli}. The most important finding of this work is that the growth law, that average cell volume is proportional to the scaling factor $S = 2^{(C+D)/\tau}$, remained valid across large perturbations in cell dimensions. As discussed in the Introduction, the growth law can be quantitatively derived under two assumptions: (i) the CH model, that replication initiation triggers cell division after a constant time $C+D$, and (ii) that the average cell volume at initiation of DNA replication is proportional to the number of origins at initiation. The robustness of the growth law as documented above suggests that both assumptions hold in face of the perturbations studied here.

\begin{figure}[t]
\centering
\includegraphics[width=3in]{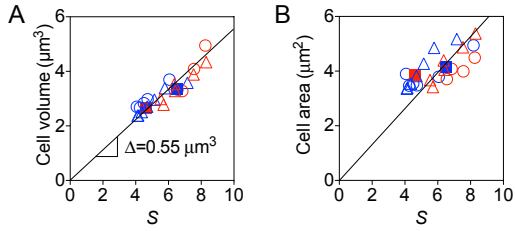}
\caption{The growth law holds in face of perturbation to cell dimensions. (A) The average cell volume is proportional to the scaling factor $S=2^{(C+D)/\tau}$. The black line shows the growth law $V=S\Delta$ for the best fit proportionality constant $\Delta = 0.55 \pm 0.04 \mu$m$^3$. The plus-minus indicates the $95\%$ confidence interval of the fit. The coefficient of determination $R^2$ of the fit is $0.81$. (B) The average cell area is not proportional to the scaling factor. The black line shows the best fit with intercept forced to zero. The $R^2$ of the fit is $0.40$. The circle, triangle, and square indicates \emph{mreB}-titratable, \emph{ftsZ}-titratable, and wild type strains, respectively. Different colors denote growth media: red is RDM+glucose, blue RDM+glycerol. The SEMs of three replicates were smaller than the size of the symbols.}
\label{Fig4}
\end{figure}

Correspondingly, given the first assumption, the presented results support the second class of models for control of cell division (Introduction) in which the timing of initiation governs that of division, and opposes the first class of models in which the timing of division is governed by the accumulation of cell size with no explicit reference to DNA replication. The presented findings do not rule out models in which, oppositely to the CH model, division triggers initiation. However, these models are heavily challenged by the finding that $C+D$ is essentially constant (CV of approximately $0.1$) on the single-cell level at fast growth \cite{elf}. To be consistent with this finding, these models must somehow coordinate the triggered initiation event with the division event in the next generation. Further studies will be required to investigate this possibility.

Our findings also provide information about the question of what key phenomenological variable governs cell cycle progression. We find that only cell volume, and not surface-area, length, or width, is proportional to the scaling factor $S$ (Fig. \ref{Fig4}, Fig. S4). Together with the second assumption (above), the robustness of the growth law then supports the hypothesis that volume per origin, rather than other geometric features, is the key phenomenological variable that is invariant at initiation. Hence, cell volume governs the timing of initiation and, by the CH model, the timing of division.

\begin{figure}[t]
\centering
\includegraphics[width=3in]{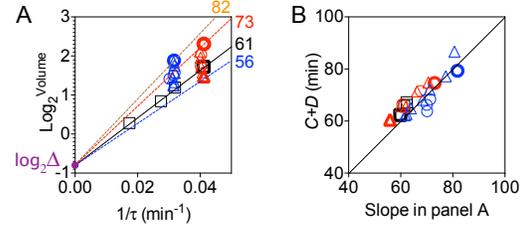}
\caption{The average cell volume per origin at initiation is constant in face of perturbations in cell dimensions. (A) Colored lines connect $\log_2(\Delta)$ to bolded symbols as examples. The respective slopes of the lines are computed as the ratio of $\log_2(V/\Delta)$ over $1/\tau$, and shown as numbers (minutes). (C) Measured $C+D$ values are plotted against the ratios calculated in panel (A). The circle, triangle, and square indicates \emph{mreB}-titratable, \emph{ftsZ}-titratable, and wild type strains, respectively. Different colors denote growth media: red is RDM+glucose, blue RDM+glycerol. The SEMs of three replicates were smaller than the size of the symbols.}
\label{Fig5}
\end{figure}

Figs. \ref{Fig4} and \ref{Fig5} also show that the proportionality constant $\Delta$, which links average cell volume to the scaling factor $S$ in the growth law (Eq. \ref{growthlaw}), remained constant across the perturbations studied here. In the derivation of the growth law, $\Delta$ is proportional to $\Delta_I$, or the average volume per origin required for initiation. The constancy of $\Delta$ therefore suggest that none of the perturbations studied (MreB, FtsZ, and various growth media) affected the molecular mechanism underlying the regulation of initiation. Despite drastic changes in cell shape, the titratable strains still initiated replication on average at a constant $0.79 \pm 0.06 \mu$m$^3$ per origin just as in wild type. This value for the average cell volume per origin at initiation is similar to the recently reported value of $0.9 - 1.0 \mu$m$^3$ \cite{elf}.

Although the present study examined altered genetic conditions that result in altered cell length and cell width, our analysis implies that these changes in cell length and cell width did not affect cell volume directly. Rather - since neither genetic perturbation affected growth rate, the $C$ period, or $\Delta$ - changes in the $D$ period alone were responsible for the observed changes in cell volume, which in turn are manifested as changes in cell length and cell width.

Fig. \ref{Fig1}B illustrates schematically how the two genetic perturbations might exert their effects. In one case, reduced \emph{ftsZ} expression increases the $D$ period. Since FtsZ is a direct mediator of septum formation, this effect could result directly from a prolongation of the septation process. The increased $D$ period dictates an increased average cell volume, which, in this situation, happens to manifest as an increase in length alone, with no change in width. This asymmetric change in length, versus width, matches the fact that cell width is maintained normally even in filamentous cells where septation is completely eliminated \cite{nhill,bending}.

Reduced \emph{mreB} expression also increases the $D$ period. In one possibility, MreB would have two direct roles: both in septation via local effects at the division site \cite{gerdes}, and in determining cell width. Here, at reduced MreB levels, delayed septation would increase the $D$ period. Due to the second role of MreB, the corresponding increase in cell volume in this case is implemented mostly by the increase in cell width. Alternatively, MreB might play only the role of determining cell width. In this scenario, the increased cell width then results in a prolonged septation process and thus, a longer $D$ period. In both scenarios, however, given experimentally-determined values for the $D$ period and cell width, the cell volume predicted by the growth law also dictates small changes in cell length. The magnitudes of these predicted changes are just at the level of detection of the current study but are consistent with observed values (Fig. S5).

In general, the growth law specifies cell volume without reference to the aspect ratio of the rod-shape morphology. As a result, a given change in cell volume due to a change in the parameters of the growth law can be manifested as diverse combinations of changes in cell length and cell width.

Our analysis has shown that, in analyzing size-related measurements, the growth law and its underlying tenets imply that any perturbations to the $C$ or $D$ periods, or growth rate will affect cell volume - even if the perturbations are not affecting the core mechanism of size regulation that determines the value of the invariant average cell volume per origin at initiation. Thus, with respect to determining cell size, an important distinction can be made between ``primary'' and ``secondary'' regulators, as highlighted in Ref. \cite{boye2003coupling}. In the context of the above analysis, MreB and FtsZ appear to be secondary regulators in \emph{E. coli} because $\Delta$ remained constant across titrated \emph{mreB} and \emph{ftsZ} levels.

All of the considerations above build on the classical works of Schaechter, CH, and Donachie, which consider population average behaviors. Donachie further proposed a single-cell interpretation of his idea in which initiation occurs in a cell when the cell reaches a constant volume per origin \cite{donachie68}. However, the proposal is not compatible with experimental single-cell measurements showing adder correlations between cell sizes at births and at divisions (see Introduction) because it predicts no such correlations under conditions where growth rate is essentially constant \cite{manalis,sattar,ho}.

As resolution to this conflict, some of us have recently proposed a phenomenological adder-per-origin model in which a constant volume is added between two rounds of initiations, rather than between two rounds of divisions \cite{AmirPRL,ho}. We have shown that this adder-per-origin model leads both to adder correlations and a constant average cell volume per origin at initiation. Furthermore, adder-per-origin can also explain rate maintenance and the growth law.

We conclude by discussing several unsolved questions. First, we have ignored single-cell fluctuations in the $C$ and $D$ periods in our discussion, but recent works show that they are important to understanding single-cell correlations \cite{ho,lagomarsino}. We have also not discussed \emph{E. coli} in slow growth conditions, but several works raise the possibility that \emph{E. coli} might behave qualitatively differently there. For one, previous works suggested that average cell volume per origin at initiation is not constant in such conditions \cite{bates2005chromosome, wold1994initiation}. Recent works also suggested that at slow growth, \emph{E. coli} does not show adder correlations \cite{elf}. In light of these observations, it will be important to further study the single-cell physiology of \emph{E. coli} under cell dimension perturbations at slow growth.

There has recently been much interest in the question of cell size homeostasis across all domains of life \cite{ginzberg2015being}, and we note that adder-per-origin may be applicable to other organisms as well. For example, it is known that the bacterium \emph{Bacillus subtilis} also exhibits the growth law and adder correlations. Repeating the experiment here in \emph{B. subtilis} may help probe the relations between cell dimensions and cell cycle timings in a Gram-positive bacterium. Adder correlations in cell volume were also found recently in budding yeast diploid daughter cells \cite{ilya}. Given the different morphology of these cells, which changes dramatically throughout the cell cycle and particularly at budding, it is plausible that cell volume, and not cell shape, is the key phenomenological variable governing cell cycle regulation also in this case.

Importantly, although our study here has suggested a coarse-grained, phenomenological model on the level of cell dimensions, it has not alluded to the molecular players involved. Several hypothetical molecular mechanisms, such as the accumulation of a threshold amount of an initiator protein per origin \cite{autorepressor} or the dilution of an inhibitor protein \cite{fantes}, were previously shown to implement molecularly the phenomenological model discussed here \cite{ho,ilya}. Yet despite decades of work, the molecular mechanism for cell cycle regulation in bacteria remains a fundamental unsolved question.

\acknow{This work was financially supported by NSFC (31471270), 973 Program (2014CB745202), 863 Program (SS2015AA020936), the Guangdong Natural Science Funds for Distinguished Young Scholar (S2013050016987), Shenzhen Peacock Team Plan (KQTD2015033ll7210153) to C.L and by a National Institutes of Health grant NIH RO1 GM025326 to N.K. AA acknowledges support from the A. P. Sloan Foundation. The authors acknowledge the reviewers, Terry Hwa and Arieh Zaritsky for insightful discussions.}

\showacknow

\end{document}